\documentclass[aps,prl,amsmath,amssymb,reprint,superscriptaddress]{revtex4-1}
\usepackage{graphicx}
\usepackage{amsmath}
\usepackage{txfonts}
\usepackage{dcolumn}
\usepackage{bm}
\usepackage{bbold}
\usepackage{color}
\usepackage{amsfonts}
\usepackage{amssymb}
\usepackage{mathrsfs}
\usepackage{braket}

\usepackage{caption}
\usepackage{subcaption}

\begin{document} 

\title{Strong Coupling Between P1 Diamond Impurity Centres and 3D Lumped Photonic Microwave Cavity}

\author{Daniel L. Creedon}
\affiliation{ARC Centre of Excellence for Engineered Quantum Systems, School of Physics, University of Western Australia, 35 Stirling Highway, Crawley WA 6009, Australia}

\author{Jean-Michel Le Floch}
\affiliation{ARC Centre of Excellence for Engineered Quantum Systems, School of Physics, University of Western Australia, 35 Stirling Highway, Crawley WA 6009, Australia}

\author{Maxim Goryachev}
\affiliation{ARC Centre of Excellence for Engineered Quantum Systems, School of Physics, University of Western Australia, 35 Stirling Highway, Crawley WA 6009, Australia}

\author{Warrick G. Farr}
\affiliation{ARC Centre of Excellence for Engineered Quantum Systems, School of Physics, University of Western Australia, 35 Stirling Highway, Crawley WA 6009, Australia}

\author{Stefania Castelletto}
\affiliation{School of Aerospace, Mechanical and Manufacturing Engineering RMIT University, Melbourne, Victoria 3000, Australia}

\author{Michael E. Tobar}
\email{michael.tobar@uwa.edu.au}
\affiliation{ARC Centre of Excellence for Engineered Quantum Systems, School of Physics, University of Western Australia, 35 Stirling Highway, Crawley WA 6009, Australia}

\date{\today}


\begin{abstract}
We report strong coupling between an ensemble of N impurity (P1) centres in diamond and microwave photons using a unique double post re-entrant cavity. The cavity is designed so that the magnetic component of the cavity field is spatially separated from the electric component and focused into the small volume in which the diamond sample is mounted.  The novelty of the structure simultaneously allows high magnetic filling factor (38.4\%) and low frequencies necessary to interact, at low magnetic field, with transitions in diamond such as those in NV$^{-}$ and P1 centres. Coupling strength (or normal-mode splitting) of 51.42 MHz, was achieved with P1 centres at 6.18 GHz and 220 mT in a centimetre-scale cavity, with a corresponding cooperativity factor of 4.7. This technique offers an alternative way, with some significant advantages, to couple 3D cavities to transitions in diamond and achieve the strong coupling necessary for applications to quantum information processing.
\end{abstract}

\maketitle

Quantum systems that consist of spin ensembles coupled to a cavity mode are attracting attention as a promising physical realization for processing and storage of quantum information, and for the realization of ultra-sensitive high resolution magnetometers\cite{Taylor2008} or to improve conventional electron spin resonance sensitivity\cite{PhysRevB.86.064514}.  A large ensemble of spins can allow collective coupling to the cavity modes, enabling the strong coupling regime to be reached\cite{Wallraff:2004lq}, - the case in which the coupling strength between a photonic cavity mode and a spin ensemble exceeds the average linewidth of these two resonances. Whereas the coupling strength to individual spin is small, the collective coupling strength $g_c$ is enhanced with a characteristic scaling factor of $\sqrt{N}$, where $N$ is the number of identical spins in the interaction volume.

The spin ensemble can be used to store quantum information if strongly coupled to the cavity modes, which can then be used as reliable readout system. Different types of spin ensembles have been studied for this purpose\cite{PhysRevLett.105.140502,PhysRevLett.105.140501,PhysRevLett.110.067004}, with recent focus (due to scalability) on solid-state systems such as color centres in diamond\cite{PhysRevLett.105.140502,Zhu2011,PhysRevLett.107.060502}, or rare-earth spin ensembles \cite{PhysRevB.84.060501,PhysRevB.90.100404}.  For the cavity subsystem, three-dimensional cavities in particular show great promise. In the case of transmon qubits, coupling to a 3D cavity has enabled qubit decoherence to be suppressed when compared to 2D stripline resonators\cite{PhysRevLett.107.240501}.  Strong coupling has also been achieved between an electron spin ensemble and a 3D microwave cavity at room temperature \cite{10.1063/1.3601930}, and between spin ensembles and several other systems\cite{Boero2013133,PhysRevLett.105.140501}. Strong coupling of nitrogen-vacancy defects in diamond to stripline resonators\cite{PhysRevLett.105.140502} is particularly interesting, as nitrogen is one of the most common impurities in diamond. 
\begin{figure}[t!]
			\includegraphics[width=0.45\textwidth]{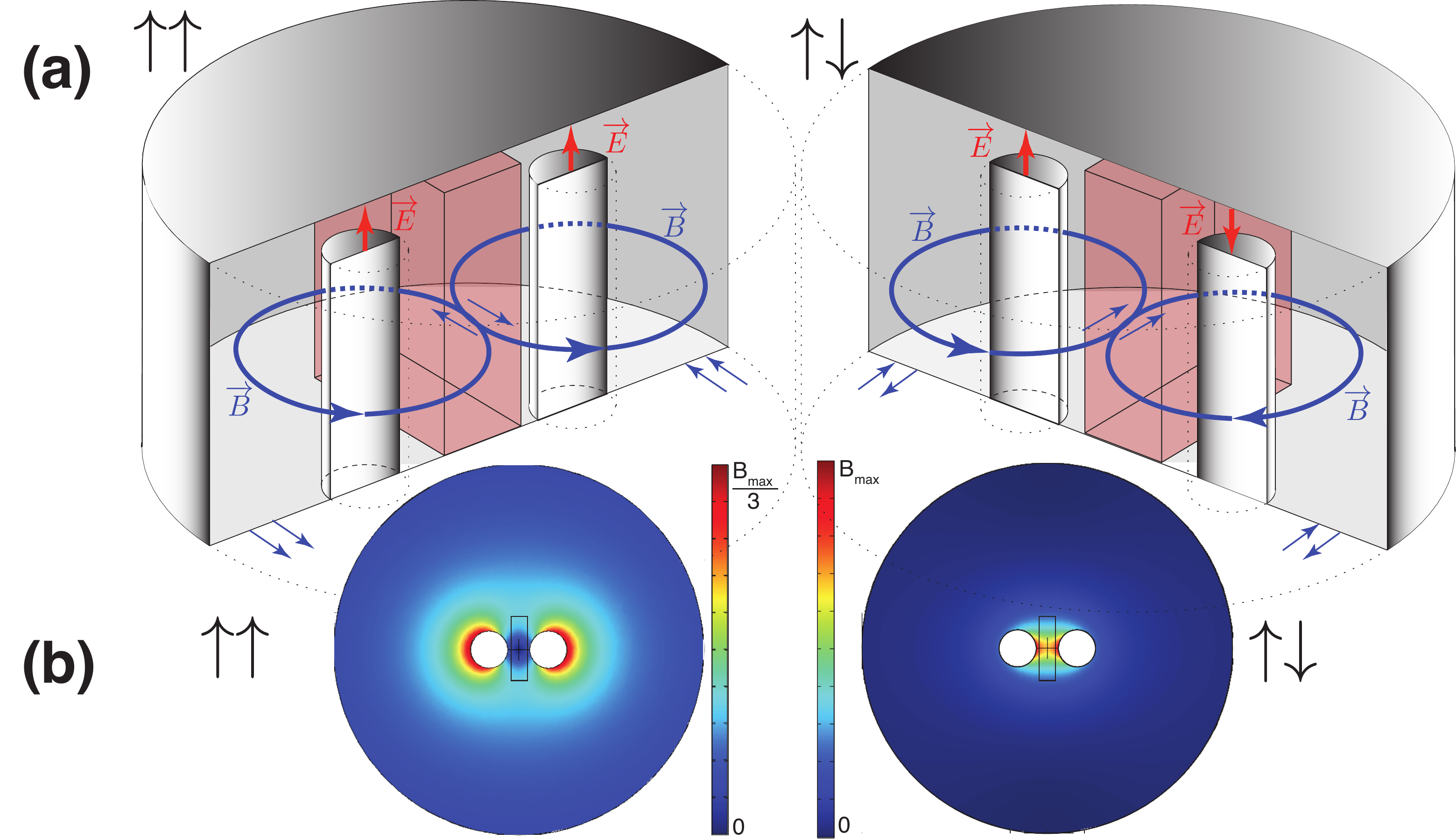}
	\caption{\label{fig-cavity}The field focusing re-entrant cavity used for the experiment. The shaded block between the posts is the diamond crystal. (a) Shows a 3D cross section of the cavity operating in the dark mode (left) and bright mode (right). (b) Shows the magnetic field density for the dark (left) and bright (right) modes in the plane at the centre of the cavity height.}
\end{figure}
Impurity spins and defect centres in diamond\cite{Loubser1978} have recently played a significant role for their potential application as qubits in the field of quantum information, computing, and control. One of the most well studied defects in diamond is the nitrogen vacancy (NV) defect due to its extremely long electron spin coherence time of 1 second, even at room temperature\cite{Bar-Gill2013}. The electron spin (S=1) of the negatively charged defect (NV$^-$) can be read-out optically via a ``flying" qubit, and can be coupled to nearby $^{13}$C nuclear spins. Coupling to other paramagnetic impurities in close proximity can also be achieved, such as the P1 impurity of diamond ($S=1/2$), commonly referred to as the C impurity - a substitutional replacement of a carbon atom in the lattice with a nitrogen atom, most commonly $^{14}$N. Such cavity coupling to nearby NV dark spins is relevant for quantum information and sensing\cite{PhysRevLett.110.157601,PhysRevLett.97.087601}, and defects in diamond whose electron or nuclear spin state in ensemble could be coherently read out can offer a large space for implementing coherent quantum control of their spin, or for use as a quantum memory. Here, we demonstrate operation in the strong coupling regime for an ensemble of P1 impurities in diamond coupled to a 3D photonic cavity mode at microwave frequencies.

\begin{figure}[t]
\centering
		\includegraphics[width=.45\textwidth]{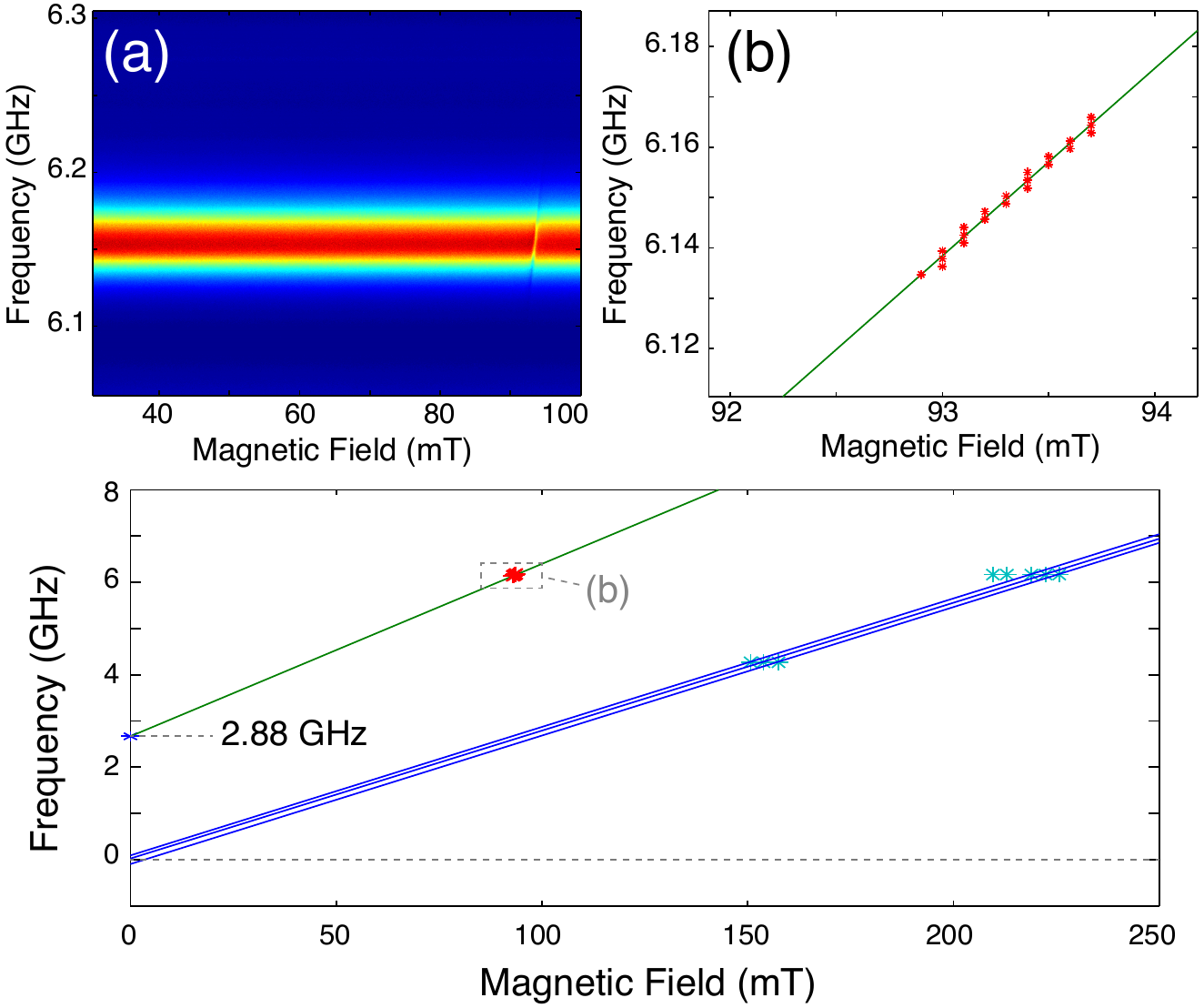}
		\caption{(a) Transmission spectrum of the `lowfield' interaction where the NV$^{-}$ transition between the $m_S =0$ and $m_S=+1$ state is tuned across the cavity BM near 93 mT. (b) Frequency at the point of minimum transmission through the cavity over the range of the NV interaction. The slope of the line confirms the well known zero-field splitting of $\sim$2.87 GHz for this defect. (c) Large span plot incorporating sub-figure (b) as well as showing interactions between nuclear hyperfine transitions of P1 centres with the cavity BM and DM.}
	\label{fig:lowfieldinteraction}
\end{figure}
Measurements were performed on two diamond crystals (herein referred to as Sample 1 and Sample 2). The diamonds are high pressure, high temperature Sumitomo type Ib bulk diamond, top and bottom surface cleaved along $\langle 100 \rangle$ faces, and irradiated with 2 MeV electrons to a fluence of $1.11\times 10^{18}/\text{cm}^2$ at $T<80^{\circ}C$, to create vacancies, and then annealed in vacuum at 900$^{\circ}C$ for 2 hours to form the NV centres.  Independent electron paramagnetic resonance spectra of the diamonds taken after electron irradiation and annealing show clearly the P1 centre region arising from substitutional N atoms and the NV centre. The nitrogen (P1) concentration was determined to be 8.2-10 parts per million, while the NV concentration was 3.1 ppm. The samples, each with dimensions $3.66 \times 3.66 \times 1.72$mm, were cleaned in nitric acid and measured separately after being mounted at the center of the dual post re-entrant cavity, shown in Fig. \ref{fig-cavity}. The principle of operation of a similar cavity has been described in detail previously, to couple magnons in a YIG sphere near 20 GHz\cite{PhysRevApplied.2.054002}. In this case we illustrate the versatility of the cavity by designing the dark and bright modes in the lower GHz range relevant for coupling to spins in diamond. The cylindrical volume of the cavity contains two cylindrical posts with height several tens of microns less than the height of the cavity. The DC magnetic field is applied in the $\langle 100 \rangle$ direction, therefore forming a $\theta=$55$^{\circ}$ angle with the 4-possible orientations of NV and P1 along the $\langle 111 \rangle$ crystallographic axes.

The two lowest order modes which may be excited in this novel cavity configuration are the so-called `dark mode' (DM) at 4.28 GHz, and the `bright mode' (BM) at 6.18GHz. The DM has a symmetric electric field component in the pair of re-entrant cavity post gaps, which leads to cancellation of the magnetic field component at the centre of the cavity due to the resulting antiparallel magnetic field contribution from each post. Thus, operation in the DM leads to a low magnetic field filling factor, i.e. the fraction of magnetic field energy inside the volume of the diamond sample is low compared to the total volume of the magnetic energy in the cavity.

Conversely, the electric field of the BM is antisymmetric in the post gaps, which leads to enhancement of the magnetic field between the posts and a subsequently large magnetic field filling factor in the volume of the crystal. This type of cavity, a so-called multiple post re-entrant cavity\cite{patent2014,multipost}, allows a large magnetic filling factor to be achieved in very small samples, something that would ordinarily require the use of a very small cavity with correspondingly (and usually undesirably) high eigenfrequencies. In the present work, given the dimensions of the diamond, cavity, and gap spacing used, we compute magnetic filling factors of 38.4\% and 2.1\% for the BM and DM respectively\cite{PhysRevApplied.2.054002}.


\begin{figure*}[t]
			\includegraphics[width=\textwidth]{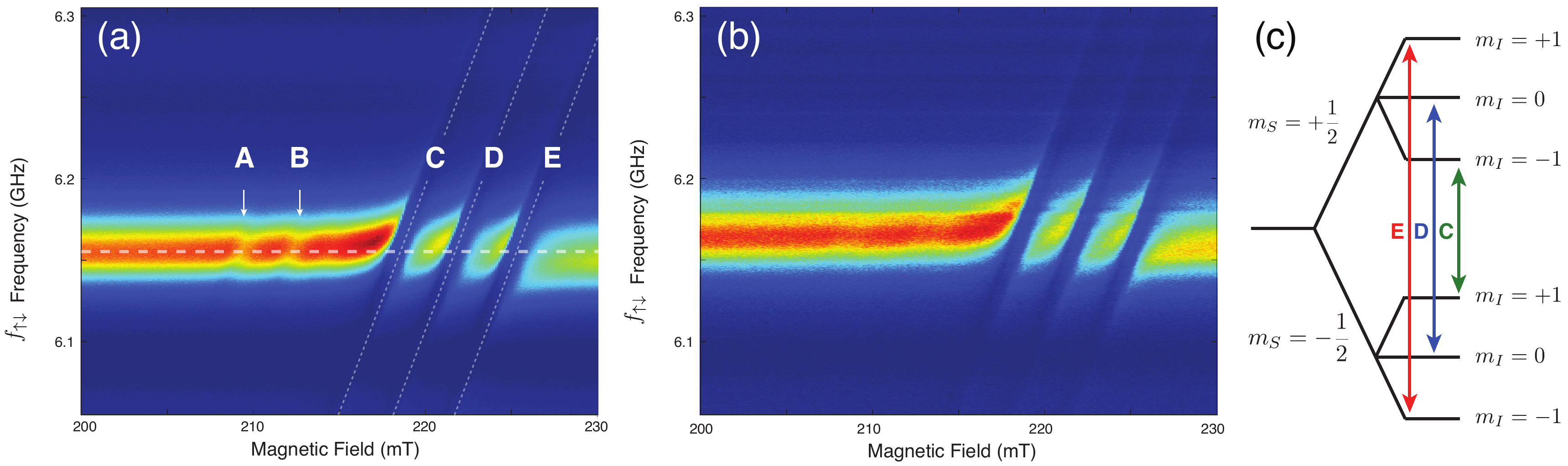}
	\caption{\label{fig1transmission}Transmission spectrum of the cavity near the bright mode with (a) Sample 1 and (b) Sample 2.  The normal behaviour of the cavity bright mode (i.e. no magnetic field dependence) is indicated in (a) as a horizontal dashed line. Three clear avoided crossings are visible (labeled C,D, and E), corresponding to the electron spin transitions of the P1 centres that conserve the $^{14}$N nuclear spin, as well as two weaker interactions (A and B) indicated by arrows. The energy level structure of the P1 centre is shown in subfigure (c).}
\end{figure*}
The measurement technique employed in this work is similar to previous works investigating spins in solids at low temperatures using dielectric\cite{PhysRevB.88.224426,PhysRevA.89.013810,Goryachev:2014aa,karim2} and metallic cavities\cite{PhysRevApplied.2.054002}.
This approach has been used previously in whispering gallery mode resonators based on crystals with low dielectric loss, as the corresponding high $Q$-factor and the existence of numerous closely spaced high order modes allows them to act as sensitive antennae for the detection of tuned electron spin resonances. Whispering gallery mode techniques have also been used previously for the characterisation of polycrystalline diamond\cite{JMdiamond}. In the case of whispering gallery mode resonators, the large population of many dozens of modes allows spin resonances to be tracked over a wide frequency range spanning several tens of gigahertz. However, in the present work it is more critical to achieve the highest magnetic filling factor in the diamond volume, and as such the accessible frequency span is closely centered around the bright mode eigenfrequency. Although higher order re-entrant cavity modes exist, the lowest order BM has the highest magnetic filling factor inside the diamond samples, and as such only this mode is monitored in the present work.

We consider the following Hamiltonian describing the interaction of NV$^-$ spin and the P1 electron and nuclear spin with a magnetic field \cite{PhysRevLett.97.087601,PhysRevB.84.193204,PhysRevB.42.8605,doi:10.1080/00268978200100921,PhysRevB.86.195316}:

\begin{multline}
H = D\left[ \left(S_z^{\text{NV}}\right)^2 - \frac{1}{3}\left(S^{\text{NV}}\right)^2 \right] +\left( S^{\text{NV}} \cdot J \cdot S^N\right) \\
+\left( S^N \cdot A \cdot I^N\right)
\end{multline}
where $S_{\text{NV}}$, is the electron spin-1 operator of NV with $z$, the NV/P1-axis direction (along 4 equivalent diamond $\langle 111\rangle$ axes), $S_N$ is the nitrogen electron spin-$\tfrac{1}{2}$ operator, and $I_N$ is the nitrogen nuclear spin-1 operator of the P1 center. The zero-field splitting of the NV electronic triplet is given by $D = 2.88$ GHz, and $J$ and $A$ are the fine-structure and hyperfine tensors respectively, with $A/2\pi=[81.33, 81.33, 114.03]$ MHz. The last terms describe the P1 centre, a nitrogen atom with an extra almost free electron interacting with the N nuclear spin and with the electronic spin of the NV$^-$ color center.


Several measurements were taken of the diamond samples in different orientations. Sample 1 was measured in two orientations, with the second being a rotation of 90$^\circ$ around the z-axis of the crystal. The rotation had a minimal effect on the interactions observed, and as such Sample 2 was measured in only 1 orientation. Using the spectroscopic technique described previously, the cavity dark mode at 4.28 GHz, and the bright mode at 6.18 GHz were examined using a vector network analyser while incrementally stepping the applied DC magnetic field. To check for power dependence, the input power to the cavity was swept over a range of 35 dB with no discernible effect other than a change in the signal to noise ratio of the measurement. Three regions of interaction were experimentally observed in the crystal. The cavity BM is seen to interact with NV$^{-}$ impurities in the diamond sample over a region of applied magnetic field near 90 mT (see Figs. \ref{fig:lowfieldinteraction}(a), \ref{fig:lowfieldinteraction}(b)), as well as P1 centres near 210-230mT (see Fig. \ref{fig1transmission}), whereas the cavity DM interacts with the P1 centres at approximately 120mT. 

Figure \ref{fig:lowfieldinteraction}(a) shows the transmission spectrum of Sample 1 as the applied magnetic field is swept and the transition frequency of the NV$^{-}$ impurity is tuned into resonance with the cavity BM, resulting in an avoided frequency crossing for an applied field near 93 mT. Figure \ref{fig:lowfieldinteraction}(b) tracks the frequency at the minimum transmission coefficient over this range of field in order to determine the slope of the interaction. This slope is extrapolated to find the zero-field splitting in Figure \ref{fig:lowfieldinteraction}(c), which assuming a constant slope was determined to be 2.87 GHz. This subfigure also shows the higher field interactions of the dark mode near 120 mT, and the bright mode near 220 mT. We attribute these interactions to the spin-flip transitions for P1 defects that conserve the nuclear spin of $^{14}$N, where $m_I =-1,0,1$. From the data, we calculate a Land\'e $g$-factor of 2.6 for the low-field NV$^{-}$ interaction (subfigure (b)), and an approximate Land\'e $g$-factor of 1.98 for the spin flip transitions (labelled C-E in Fig. \ref{fig1transmission}).

Figure \ref{fig1transmission} shows the transmission spectrum of the cavity BM when loaded with two different diamond samples. The nuclear spin of the P1 defect, most commonly $^{14}$N is $I=1$, leads to an energy level splitting into $2I+1=3$ energy levels. This hyperfine spectrum of the P1 centres is clearly visible as three avoided crossings that correspond to the allowed nuclear spin conserving electronic transitions (i.e the spin flip transitions$m_S=\pm\frac{1}{2}$, $m_I=-1,0,+1$ levels, labelled C-E respectively). Also faintly visible for both samples is a set of two weaker interactions (indicated with arrows) in the interaction with the bright mode.  This may be attributed to anisotropy in the hyperfine coupling caused by a geometrical Jahn-Teller distortion of the bond between the P1 centre nitrogen atom and a neighbouring carbon atom. The distortion is randomly oriented along one of the four $\langle 111 \rangle$ crystallographic directions, and the hyperfine splitting is determined by the angle between this distorted bond direction and the applied external magnetic field. When the applied field is parallel to the $\langle 100 \rangle$ crystallographic direction of the diamond, all of the carbon bond angles relative to the field are equivalent, and so only three hyperfine lines are seen\cite{PhysRevLett.105.140501}. However, when the field is applied along the crystallographic $\langle 111 \rangle$ direction, a set of 5 lines is observed in the spectrum\cite{PhysRevLett.110.157601,PhysRevLett.101.047601,PhysRev.115.1546}. Thus, we attribute the detection of these weak transitions to a small misalignment of the crystallographic axes of the sample within the field, or some internal strain of the crystal. Only the strongest three transitions are observed to interact with the dark mode due to its low magnetic filling factor and corresponding reduced sensitivity.

\begin{figure}[t]
	\centering
		\includegraphics[width=.45\textwidth]{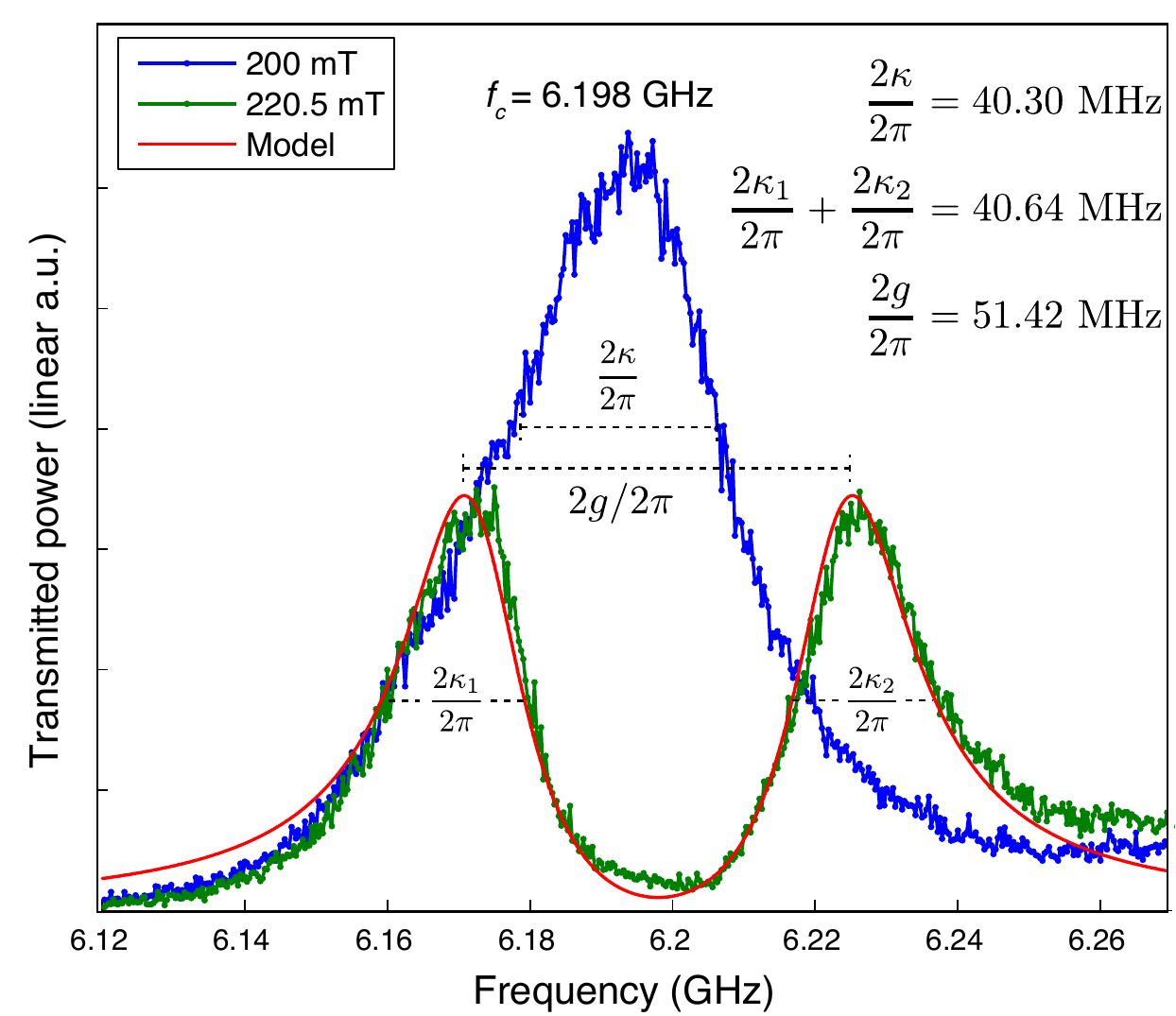}\caption{Demonstration of strong coupling between diamond P1 energy levels in Sample 1 and the bright cavity mode. Slices through Figure \ref{fig1transmission}(a) at 220 mT and 220.5 mT are shown. The slices show the BM far from the interaction, as well as on resonance where it interacts with the with the spin-flip transition labelled C in Fig. \ref{fig1transmission}. The data is well modelled by a pair of coupled oscillators. The plot shows the largest ratio of splitting to linewidth achieved.}
	\label{fig:strongcoupling}
\end{figure}

The hyperfine structure of P1 centres in diamond has been observed previously by coupling to superconducting transmission line cavities\cite{PhysRevLett.105.140501}.  Here, we demonstrate that the strong coupling regime can be achieved between P1 impurity centres in diamond and photons in a 3D re-entrant cavity due to the strong enhancement in magnetic filling factor.  Figure \ref{fig:strongcoupling} shows two slices through the transmission spectrum of the cavity BM when loaded with Sample 1 shown in Fig. \ref{fig1transmission}(a). The first slice at an applied field of 200 mT shows the normal transmission curve of the cavity bright mode far from the interaction, i.e. the bare mode centred at 6.198 GHz. In this case, the $Q$-factor of the cavity mode is approximately $Q_L=150$ when loaded with the diamond sample.  The second curve shows the mode splitting when the ($m_s=\pm\frac{1}{2}$, $m_I=-1$) transition of the P1 defect is tuned into resonance with the cavity at 220.5 mT, a splitting which is well modelled by a pair of coupled oscillators. We observe normal mode splitting of $\Delta=2 g/2\pi=51.42$ MHz,  with the sum of the split mode linewidths $2\kappa_1/2\pi$ and $2\kappa_2/2\pi$ equal to 40.64 MHz, i.e. a ratio of 1.3.  We compute from the data that the density of coupled spins is approximately $1.6\times 10^6/\mu \text{m}^3$, and deduce from the fit a dipolar spin broadening of $2\Gamma/2\pi \sim$7 MHz. Thus we calculate a cooperativity factor $C=g^2/\kappa\Gamma$ of 4.7, exceeding that measured by Schuter et al.\cite{PhysRevLett.105.140501} by nearly a factor of three.


In summary, strong coupling has been demonstrated between the nuclear spin of P1 impurity centers in diamond and the bright mode of a re-entrant magnetic field focusing 3D cavity\cite{PhysRevApplied.2.054002}. This result represents an improvement on the strong coupling condition compared to previous work with planar superconducting cavities. In addition, we observed the NV$^-$ electron spin transition and the P1 spin flip transition using the dark mode of the cavity, with comparatively low magnetic filling factor. This work opens new avenues for the use of high magnetic filling factor cavities to achieve the strong coupling regime with greater collective spin coupling than previously observed, and with a variety of spin qubits. 
The novel patented cavity design\cite{patent2014,multipost} can be engineered to yield an arbitrary spectrum of cavity modes, and in principle can be designed in such a way that the bright mode frequency corresponds to the zero-field splitting of the NV defect, which enables the use of superconducting cavities and qubits by virtue of the fact that an applied DC magnetic field is no longer required. It may also be designed to allow the application of large external magnetic fields which can be beneficial in terms of reducing spin-bath decoherence. Finally and more generally, the system represents a novel and sensitive method for electron paramagnetic resonance studies in solids. 

\begin{acknowledgments}
The authors wish to acknowledge that this work was supported by Australian Research Council grant CE110001013, and a University of Western Australia Research Collaboration Award. We also thank Brett Johnson and Takeshi Ohshima at JAEA for the diamond irradiation.
\end{acknowledgments}

\end{document}